# DØ Regional Analysis Center Concepts

L. Lueking, representing the DØ Remote Analysis Task Force
*FNAL, Batavia, IL 60510*

The DØ experiment is facing many exciting challenges providing a computing environment for its worldwide collaboration. Transparent access to data for processing and analysis has been enabled through deployment of its SAM system to collaborating sites and additional functionality will be provided soon with SAMGrid components. In order to maximize access to global storage, computational and intellectual resources, and to enable the system to scale to the large demands soon to be realized, several strategic sites have been identified as Regional Analysis Centers (RAC's). These sites play an expanded role within the system. The philosophy and function of these centers is discussed and details of their composition and operation are outlined. The plan for future additional centers is also addressed.

## 1. INTRODUCTION

The DØ experiment is facing many computing challenges on the journey to achieving our physics mission. There are billions of recorded triggers, and dozens of physics analysis areas. Theses analyses are complex, precision measurements with minute signals and/or subtle systematic errors. Some examples include: understanding the underlying event consistent with 5 MeV/c$^2$ statistical precision on $M_W$, determining the jet energy scale to more precisely measure $M_{top}$, and tagging the vertex of $B$ mesons in an environment of 5-10 overlapping interactions. We estimate that at the end of the current data taking period, referred to as Run 2a (2002 through mid-2004), the computing needs for MC, Reconstruction, and Analysis will be of order 4 THz CPU, and 1.5 PB storage. The needs beyond 2004 are larger still with data storage estimated to increase by 1 PB per year.

There are many potential resources available, but there are problems utilizing them. Technology and computing hardware abound, CPU and memory are inexpensive, networking is becoming more pervasive, disk and tape storage is affordable. An army of Physicists, Over 600 collaborators are "available" to manage the hardware and perform the needed processing. But, these resources are not all in one place, and they are not really "ours". The resources are distributed around the world at 80 institutions in 18 countries on 4 continents. In most places, the resources are shared with other experiments or organizations. Management, training, logistics, coordination, planning, estimating needs, and operation are difficult. The infrastructure and tools needed to pull this all together are essential.

Our strategy, to achieve our goals and best utilize available resources, is to divide and conquer. We plan to accomplish this by identifying six to ten geographical/political regions. Within each region we will establish a Regional Analysis Center (RAC), and define the responsibilities for each region. This will enable the effective use of all resources; hardware, informational, and human. We are in the process of laying the basic infrastructure now, and we will fine-tune it as our understanding of the needs evolve.

## 2. REMOTE AND REGIONAL COMPUTING

DØ has had a vision of enabling distributed computing resources for several years. The history of the process includes many important steps along the way:

- 1998: DØ Computing Model- The distributed computing concepts in SAM [1,2] were embraced by the DØ management. All of DØ 's Monte Carlo was produced at remote centers.
- 2001: D0RACE – Remote Analysis Coordination Effort [3] team helped to get the basic DØ infrastructure to the institutions. With this effort, 60% of the DØ sites have official analysis code distributions and 50% have SAM stations.
- 2002: RAC grassroots team – Met throughout spring and summer to write a formal document outlining the concepts [4].
- 2002: OATF - Offsite Analysis Task Force – Charged by the Spokespersons to further study the needs of offsite computing and analysis
- 2003: DØ Finance committee – decides how the collaboration as a whole will contribute remote computing resources to the experiment.
- 2003: Plans for MOU's are being made.

### 2.1. The Importance of Regions

Establishing regional computing will provide many important computing related advantages and opportunities for the collaboration. Opportunistic use of all computing resources within the region will be enabled. Management for resources within the region will be provided. Coordination of all processing efforts is easier through a well defined command structure. Security issues within each region are likely to be similar, as they will share Grid Certificate Authorities, security policies, and so on. An increase in the overall technical support base is anticipated as experts are identified within each region.

TUAT002

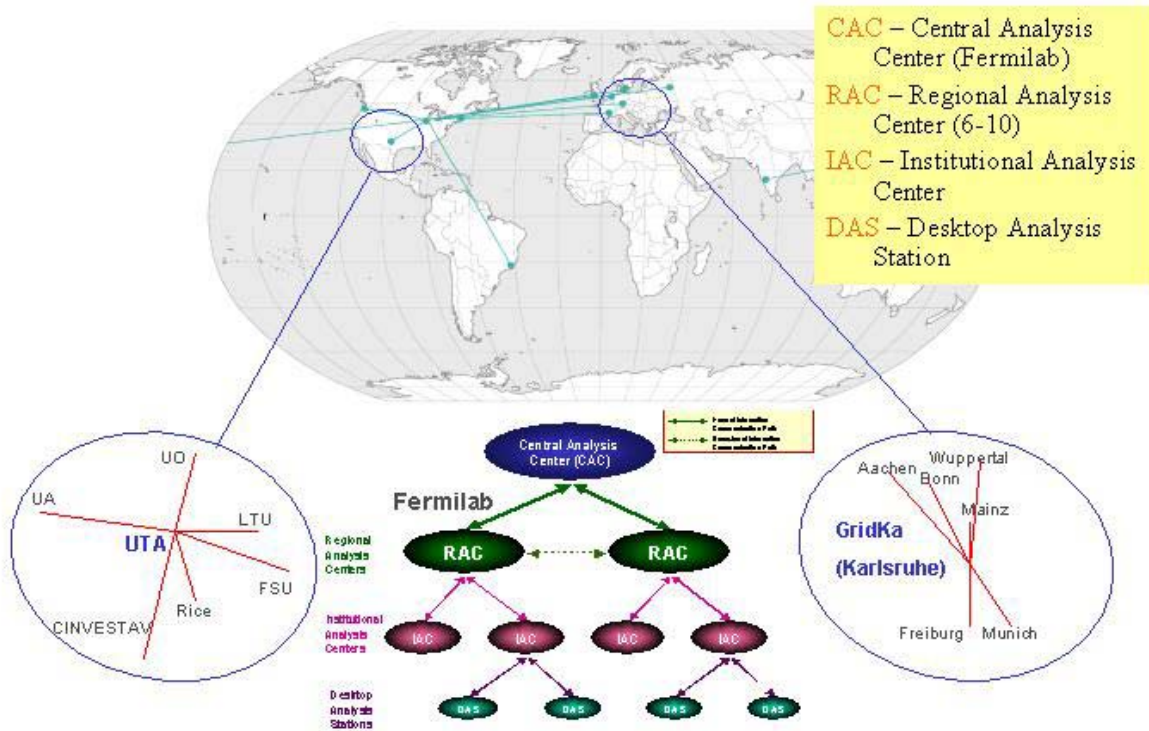

Figure 1: The hierarchical model showing examples of DØ regions being defined for the Southern US, and Germany, with their Regional Analysis Centers at UTA and GridKa.  Regions are also defined in France, The UK, and the Northern part of the US. The Insets show the model elements with Central, Regional, and Institutional Analysis Centers.  Desktop Analysis Stations are the lowest tier.

Communication within each region will be optimized as members will speak the same language and share the same time zone. Frequent Face-to-face meetings are essential and will be natural among participants within each region. Physics collaboration at a regional level will contribute to results for the experiment at large. Of course, a little spirited competition among regions is good to spur the acquisition and most effective use of resources.

### 2.2. Hierarchical Model

From these regional concepts is emerging a new way to deploy the system and manage and support the flow of data and processing. DØ has a large install base of software already in place with the majority of the institutions having the DØ code and data handling installations in place.  This initial deployment was accomplished using a Fermi-centric model with most of the support done by experts at FNAL. In addition, the hub of the data-intensive operation has been Fermilab with most SAM stations configured to send produced data to, and retrieve needed data from, FNAL. At least initially, the regional centers are at the heart of a hierarchical deployment plan and they will represent focal points for the computing operations in the coming months.  This model is shown in Figure 1 with RAC's indicated at UTA and GridKa, and their associated institutional participants.

### 3. REGIONAL CENTER FUNCTIONALITY

The functionality of the RAC will be many fold and we have made initial estimates of how we may partition the resources for various tasks. Pre-emptive caching of data will be coordinated globally so that all DST's will be on disk at the sum of all RAC's. All Thumbnail (TMB) files will be on disk at each RAC to support the mining needs of the region. In addition there will be data which is coordinated regionally to include other formats such as derived data and Monte Carlo data.  In addition, some fraction, 10% or more, of the disk will be used as on-demand SAM cache.  In addition, each center will provide some archival tape storage for selected MC samples and secondary data sets as needed by the region, or the collaboration in general. Table 1 summarizes the DØ Data Model, with multiple tiers representing output data from the analysis chains for the detector and Monte Carlo data. The event sizes for each tier are indicated, and the proposed percentage of data stored at FNAL and at each RAC, on tape and disk, are shown.

Table 1: The DØ data model, and possible percentages of each tier of data to be stored at Fermilab and at regional analysis centers on disk and tape. The bottom row indicates our estimate of the total storage required at Fermilab and the regional centers for Run 2a data processing.

| Data Tier | Size/Event (kB) | FNAL Tape | FNAL Disk | RAC Tape | RAC Disk |
|---|---|---|---|---|---|
| RAW | 250 | 100% | 10% | 0 | 0 |
| Reconstructed | 500 | 100% | 10% | 1% | 0 |
| DST | 150 | 100% | 100% | 10% | 10% |
| Thumbnail | 10 | 400% | 100% | 100% | 100% |
| Derived data | 10 | 400% | 100% | 100% | 100% |
| MC D0GSTAR | 700 | 0 | 0 | 0 | 0 |
| MC D0SIM | 300 | 0 | 0 | 0 | 0 |
| MC DST | 400 | 100% | Few % | Few % | Few % |
| MC TMB | 20 | 100% | 100% | 0 | 10% |
| MC PMCS | 20 | 100% | 100% | 0 | 10% |
| MC root-tuple | 20 | 100% | 0 | 10% | 0 |
| Totals for Run 2a (2001-mid 2004) | | 1.5 PB | 60 TB | ~50 TB | ~50 TB |

Each regional center will need CPU capable of supporting work both within the region and for the collaboration at large. This computing power will support analysis, data re-reconstruction, MC production and general purpose DØ processing needs. Also required is sufficient network throughput to support intra-region, Fermilab to region, and inter-RAC connectivity.

### 3.1. Requirements

Each center will be required to install and maintain a set of minimal infrastructure components. These servers are required to be operated on one, or a few, gateway nodes which have specific requirements for network access and software products. These items include the SAMGrid[5] servers that enable grid access to the overall DØ Grid system for data handling and computing job and information management. In addition, for certain kinds of processing activity access to database information maintained at Fermilab, such as detector calibration, is required. Proxy servers, called Database Analysis Network (DAN)[6], will be maintained at each center to provide this functionality for each region. The system is designed to accommodate the policies and culture of each center, such as firewalls, workers on private networks, and sharing compute resources with other experiments.

### 4. CHALLENGES AND PROSPECTS

As DØ deploys the components we face many interesting challenges. Operation and Support is currently provided for the SAM system through a "helpdesk" type operation manned by shifters made up of trained physicists. This support is supplied on a 24/7 basis by enlisting personnel distributed among the many time zones represented by the DØ collaboration. SAMGrid station administrators are trained and much of their expertise is based on experience installing and maintaining the system.

We have established a Grid Technical Team composed of experts in SAM-Grid development team, core SAM team, and DØ software and technical experts from each RAC. The hardware and system support is provided by the centers. Production certification is an important and time consuming part of the operation with all DØ MC, reconstruction, and analysis code releases requiring this crucial step. Special requirements for certain RAC's forces customization of infrastructure and introduces deployment delays. Security issues represent a major concern with details regarding grid certificates, firewalls, and site policies being negotiated.

### 4.1. Progress and Prospects

The first RAC prototype was chosen to be tested at GridKa in Germany and this has provided valuable experience for the future program. This center is located at Forschungszentrum Karlsruhe (FZK) and the regional participants include Aachen, Bonn, Freiburg, Mainz, Munich, and Wuppertal. GridKa has been identified in Germany as a Regional Grid development, data and computing center. The facility was established 2002 serving eight HEP experiments: Alice, Atlas, BaBar, CDF, CMS, Compass, DØ, and LHCb.

Table 2: Regional resources currently identified. Numbers in () represent the total resources available at each center, and resources allocated to DØ are indicated by the other numbers.

| RAC | IAC's in Region | CPU $\Sigma$Hz (Total*) | Disk (Total*) | Tape (Total*) | Schedule |
|---|---|---|---|---|---|
| GridKa @FZK | Aachen, Bonn, Freiburg, Mainz, Munich, Wuppertal, | 52 GHz (518 GHz) | 5.2 TB (50 TB) | 10 TB (100 TB) | Eatablished RAC |
| SAR @UTA (Southern US) | AZ, Cinvestav (Mexico City), LA Tech, Oklahoma, Rice, KU, KSU | 160 (320 GHz) | 25 TB (50 TB) |  | Active MC production center. Computing in this table available Summer 2003 |
| UK @ TBD | Lancaster, Manchester, Imperial College, RAL | 46 GHz (556 GHz) | 14 TB (170 TB) | 44 TB | Active, MC production. RAC functionality later this year. |
| IN2P3 @Lyon | CCin2p3, CEA-Saclay, CPPM-Marseille, IPNL-Lyon, IRES-Strasbourg, ISN-Grenoble, LAL-Orsay, LPNHE-Paris | 100 GHz | 12 TB | 200 TB | Active, MC production. RAC functionality later this year. |
| DØ@FNAL (Northern US) | Farm, cab, clued0, Central-analysis | 1800 GHz | 25 TB | 1 PB | Established as CAC |

The GridKa DØ RAC has been automatically caching thumbnail data produced at Fermilab since the summer of 2002. Production certification was accomplished by analyzing identical data samples at FNAL and GridKa, and comparing the results. The center provided computing resources used to produce results presented at the Winter conferences. Some Monte Carlo production was performed there, and the resources have been very effectively used by DØ.

Several additional regional centers are being constructed as summarized in Table 2. The potential for additional CPU and caching disk is large although we are just beginning to forge formal agreements with resource suppliers concerning DØ allocations. The sum of allocations at remote sites, as it is understood now, amounts to about 360 GHz. The total CPU resources at currently identified remote sites is over 1800 GHz. This compares to the 1800 GHz at Fermilab, and the need for over 4 THz by the end of the Run 2a period in mid 2004. There additional CPU and storage at institutional centers which have not been accounted for here. Nevertheless, although we are still short of our projected need, we feel that the DØ computing performed at regional centers, and within regions will soon meet a significant fraction of the experiment's needs, possibly even larger than the FNAL contribution by mid 2004.

## 5. CONCLUSIONS AND PLANS

DØ has had a distributed computing vision for many years and has built a computing infrastructure to support this. We feel that the Regional Analysis Center approach is important to more effectively use remote resources including hardware, support, and intellectual. Management and organization in each region is as important as the hardware.

In spite of the enthusiasm for this approach there are still many lessons to learn. It is understood that physics group collaboration will transcend regional boundaries outlined in this plan. In the Grid computing model, resources within each region will be used by the experiment at large. Our models of usage will be revisited frequently as we better understand the usage patterns, and experience already indicates that the use of thumbnails differs from that of our RAC model. We understand that no RAC will be completely formed at birth, and each facility will grow and evolve to meet the needs of the region, the DØ experiment, and the HEP community. There are many challenges ahead and we continue to learn and cultivate the regional center concept.

## Acknowledgments

We acknowledge all those in DØ who have diligently worked to establish these ideas and have worked hard to advance them. The members of the Off Site Analysis Task Force, official and non-official, who have contributed to the final draft of the recommendations to the DØ spokespersons are listed below with their institutional affiliations.

Iain Bertram – Lancaster University, UK
Chip Brock, Dugan ONeil – Michigan State University
John Butler – Boston University
Gavin Davies, Rod Walker – Imperial College, United Kingdom
Amber Boehnlein, David Fagan, Alan Jonckheere, Lee Lueking, Don Petravik, Vicky White (co-chair) -Fermilab
Nick Hadley (co-chair) - University of Maryland
Sijbrand de Jong - University of Nijmegen, The Netherlands
Peter Maettig, Daniel Wicke, Christian Schmitt - Wuppertal, Germany
Christian Zeitnitz – Mainz, Germany
Pierre Petroff (co-chair) - Laboratoire de l Accelerateur Lineaire, France
Patrice Lebrun – ccin2p3 in Lyon, France
Jianming Qian – University of Michigan
Jae Yu – University of Texas Arlington

We thank those who have made these visions a reality through DØRACE deployment efforts, and the SAMGrid project. We would like to thank all of those who have contributed their time and effort to making the RAC concept work at GridKa, and to those who are planning and deploying at other identified sites. At GridKa, the political structure includes Peter Mattig (Wuppertal) as the Fermilab representative to the GridKa Overview Board. Christian Zeitnitz (Mainz), and Daniel Wicke (Wuppertal) are DØ representatives to the GridKa Technical Advisory Board.